\documentclass[12pt]{article}
\usepackage[a4paper,tmargin=2cm,bmargin=2cm,lmargin=2cm,rmargin=2cm]{geometry}
\usepackage[T1]{fontenc}
\usepackage[utf8]{inputenc}
\usepackage{amssymb}
\usepackage{amsmath}
\usepackage{times}
\usepackage{color}

\usepackage[english]{babel}
\usepackage[colorlinks=true,linkcolor=blue,urlcolor=blue,citecolor=blue]{hyperref}

\usepackage{graphicx}

\newcommand{\N}{\mathbb{N}}
\newcommand{\R}{\mathbb{R}}

\newcommand{\dt}{{\rm d}t}

\newcommand{\bea}{\begin{eqnarray}}
\newcommand{\eea}{\end{eqnarray}}
\newcommand{\be}{\begin{equation}}
\newcommand{\ee}{\end{equation}}
\newcommand{\Eq}[1]{Eq.~(\ref{#1})}

\newcommand{\eq}[1]{(\ref{#1})}
\newcommand{\rme}{\mathrm{e}}
\newcommand{\rmd}{\mathrm{d}}
\newcommand{\nn}{\nonumber}
\newcommand{\ca}[1]{\mathcal{#1}}

\newcommand{\fig}[2]{\includegraphics[width=#1\columnwidth]{./#2}}

\begin{document}

\title{\bf\Large Anchored advected interfaces, Oslo model, and roughness at depinning}
\author{\bf\normalsize Assaf Shapira${}^1$ and Kay J\"org Wiese${}^2$}
\date{\small${}^1$Universit\'e Paris Cit\'e, CNRS, MAP5, 75006 Paris, France\\${}^2$CNRS-Laboratoire de Physique de l'Ecole Normale Sup\'erieure, PSL Research University, Sorbonne Universit\'e, Universit\'e Paris Cit\'e, 24 rue Lhomond, 75005 Paris, France}

\maketitle

\begin{abstract}
There is a plethora of 1-dimensional advected systems with an absorbing boundary: the Toom model of anchored interfaces,  the directed exclusion process where in addition to diffusion particles and holes can jump over their right neighbor,  simple diffusion with advection, and   Oslo sandpiles. All these models share a roughness exponent of $\zeta=1/4$, while the dynamic exponent $z$ varies, depending on the observable. We show that for the first three models  $z=1$, $z=2$, and  $z=1/2$ are realized, depending on the observable.
The Oslo model is apart with a conjectured dynamic exponent of $z=10/7$. 
Since the height in the latter is the gradient of the position of a disordered elastic string, this shows that $\zeta =5/4$ for a driven elastic string at depinning. 
\end{abstract}

\section{Introduction}
Interfaces subject to quenched disorder describe a variety of physical phenomena \cite{Kardar1997,Wiese2021}, such as domain walls in magnets \cite{DurinZapperi2006b,DurinBohnCorreaSommerDoussalWiese2016,terBurgBohnDurinSommerWiese2021}, contact-line depinning \cite{LeDoussalWieseMoulinetRolley2009},   fracture \cite{RamanathanErtasFisher1997,PonsonBonamyBouchaud2006}, or earthquakes \cite{DSFisher1998}. 
Two universality classes have to be distinguished: equilibrium and depinning.
Though many numerical studies exist \cite{RossoKrauth2001b,FerreroBustingorryKolton2012,GrassbergerDharMohanty2016}, and a field theory was developed \cite{LeDoussalWieseChauve2002,ChauveLeDoussalWiese2000a,Wiese2021}, there are   few exact results.  

A notable exception in equilibrium is the roughness exponent $\zeta_{\rm RB}^{d=1}=2/3$ for a $1+1$ dimensional directed polymer in random-bond (RB) disorder, itself related to the KPZ universality class \cite{KPZ,Kardar1997} with roughness $1/2$ and dynamic exponent $z=1/\zeta_{\rm RB}^{d=1}=3/2$ \cite{Kardar1987}.
For random-field (RF) disorder of a $d$-dimensional interface in $d+1$ dimensions in equilibrium, scaling arguments correctly predict $\zeta_{\rm RF}^d=(4-d)/3$, which can experimentally be seen even in dimension $d=0$ \cite{terBurgRissoneRicoPastoRitortWiese2023}, where it reduces to the Sinai model \cite{Sinai1983}.

At depinning,  there is  the single  random-field universality class, and no analytic result is known, apart from $d=0$ for which $\zeta_{\rm dep}^{d=0}=2^-$ \cite{LeDoussalWiese2008a}.
Based on numerical simulations, it was recently conjectured \cite{GrassbergerDharMohanty2016} that
a driven 1-dimensional string has a roughness exponent of $\zeta_{\rm dep}^{d=1}=5/4$.
Here we aim at demonstrating this value. A key observation is that if the roughness of the string at depinning  is $\zeta_{\rm dep}^{d=1}=5/4$, the Oslo model \cite{Frette1993,ChristensenCorralFretteFederJossang1996}, of which the height can be viewed as the gradient of the position of a driven string pulled at one end, has a roughness of $\zeta_{\rm Oslo}^{d=1}=1/4$.

There are   several 1-dimensional systems   with a scaling exponent  of  $1/4$, but this is generally in the temporal domain. E.g.\ a marked monomer on a polymer grows with time as $\left<[h(0,t)-h(0,t')]^2\right>\sim |t-t'|^{2H}$, with $H=1/4$. This is obtained from standard scaling arguments as $H=\zeta/z$, with $\zeta=1/2$ and $z=2$.  If these systems are advected, i.e.\ $z=1$, the roughness exponent becomes $\zeta=Hz \to 1/4$ \cite{DerridaLebowitzSpeerSpohn1991,KrugSocolarGrinstein1992,KrugSocolar1992,Krug1997,Pruessner2004}.

Here we consider a class of models which allow for a hydrodynamic description equivalent to diffusion of a scalar field $h(x,t)$ combined with advection away from an absorbing boundary at $x=0$:
\be\label{diffusion-equation}\tag{aaEW}
\begin{split}
\partial_t h(x,t) &= D \nabla^2 h(x,t) - \mu \nabla h(x,t) + \sigma \xi(x,t),\\
h(0,t) &= 0, \quad \left<\xi(x,t) \xi(x',t')\right> = \delta(x-x')\delta(t-t').
\end{split}
\ee
We call it the {\em anchored advected Edwards-Wilkinson} equation (aaEW).
While in the bulk the stationary state $h(x,\infty)$ can be obtained from a Brownian motion in the comoving frame, close to the boundary we establish
in section \ref{s:Analytic result for the variance of the height} that $\left<h(x,\infty) ^2\right>\sim \sqrt {x}$, thus $\zeta=1/4$.

This system possesses three distinct dynamical scaling behaviours, characterized by a dynamical exponent $z=1$ due to the advection, $z=2$ for bulk properties in the comoving frame, and $z=1/2$ for the decay of the equal-point correlation function (in the steady state),
\be
\left< h(x,t) h(x,0) \right> = \sqrt{\frac{x}{2\pi}} f(t/x^{1/2}). 
\ee
The function $f$ is obtained analytically in section \ref{s:Decay of the auto-correlation function}.

The model \eq{diffusion-equation}  is the proper thermodynamic description for a variety of systems. We give   two examples -- the directed exclusion process (section \ref{s:The Directed Exclusion Process (DEP)}) and the Toom interface model (section \ref{s:Toom's interface model}).
For the Oslo model (section \ref{s:Oslo model}) a similar mapping can be constructed, with a crucial difference: The {\em local time} which sets the clock for an update is advanced by the toppling itself, leading to a different dynamical exponent, conjectured to be   $z=10/7$ \cite{GrassbergerDharMohanty2016}, thus larger than the corresponding exponent $z=1$ in aaEW, but smaller than the one for diffusion ($z=2$). Still, the roughness exponent equals $\zeta=1/4$, independent of whether one considers bulk or boundary driving. The Oslo model (section \ref{s:Oslo model}) achieves this by propagating out from the seed {\em in both directions}; the seed position effectively acts as an absorbing boundary.

In the derivation of a hydrodynamic description   as in \Eq{diffusion-equation},   additional   terms may appear, 
\be\label{KPZ-equation}
 \begin{split}
\partial_t h(x,t) &= D \nabla^2 h(x,t) - \mu \nabla h(x,t) + \sum_{n\ge 2} \lambda_n \left[ \nabla h(x,t)\right]^2+ \sigma \xi(x,t),\\
h(t,0) &= 0, \quad \left<\xi(x,t) \xi(x',t')\right> = \delta(x-x')\delta(t-t').
\end{split}
\ee
The perturbation with  $\lambda_2$ is the most relevant one, leading to the   {\em anchored advected Kardar-Parisi-Zhang} universality class (aaKPZ) discussed in section \ref{s:Asymmetric case -- the aaKPZ universality class}. If this term is forbidden by symmetry, the next relevant one   is $\lambda_3$. It is marginally relevant, and believed to modify the effective parameters in the symmetric case \cite{Krug1997}, but it cannot change the stationary measure in the bulk (section \ref{s:Protection of the stationary measure}). 

This paper is organized as follows: In section \ref{s:Models in the aaEW and aaKPZ universality classes} we introduce models in the 
aaEW and aaKPZ universality classes, first the directed exclusion process (section \ref{s:The Directed Exclusion Process (DEP)}), then   
Toom's interface model (section \ref{s:Toom's interface model}); their hydrodynamic description  is obtained in section \ref{s:Continuum theory}.
In section \ref{s:Symmetric case -- the aaEW universality class} we derive analytic results for the aaEW universality class, followed by results for the aaKPZ class in section \ref{s:Asymmetric case -- the aaKPZ universality class}.
The  Oslo model is treated  in section \ref{s:Oslo model}.

\section{Models in the aaEW and aaKPZ universality classes}
\label{s:Models in the aaEW and aaKPZ universality classes}
\subsection{The Directed Exclusion Process (DEP)}
\label{s:The Directed Exclusion Process (DEP)}
In this model, each site $x = 1,2,\dots$ is either empty or occupied. Particles jump with rate $1$ to an empty neighbor as in the simple exclusion process, but an additional \emph{directed} transition is introduced, which allows particles to jump over a particle to its right with rate $1$, and holes over a hole to the right with rate $r$. The transitions are 
\begin{enumerate}
\item 10 $\to$ 01 with rate $1$,
\item 01 $\to$ 10 with rate $1$,
\item 110 $\to$ 011 with rate $1$,
\item 001 $\to$ 100 with rate $r$.
\end{enumerate}
A rate $r=1$ preserves the particle-hole symmetry, while breaking the directional (left-right) symmetry. We refer to this model as the \emph{symmetric} DEP (sDEP). The case   $r \neq 1$ is referred to as the \emph{asymmetric} DEP (aDEP), which breaks \emph{both} directional and particle-hole symmetry.

When using periodic boundary conditions, the  product measure is stationary. This can be derived from the  (non-detailed) balance equation.  Then the total particle current $j(\rho)$  is given by
\be\label{j(rho)}
 j(\rho) = 2 \left[\rho^2(1-\rho) - r (1-\rho)^2 \rho \right]= 2\rho(1-\rho)\left[ \rho - r(1-\rho)\right].
\ee
Therefore,   $\rho_{\rm c}=\frac{r}{1+r}$ is the critical density, where the overall current vanishes.

In contrast, we consider  a large system with   closed boundaries, where no particle current is   allowed. This leads to   \emph{self-organized criticality}, where particle (or hole) excess will be pushed away to the right. Taking $L\to \infty$ results in a stationary density  $\rho = \rho_{\rm c} =\frac{r}{1+r}$ in the domain of observation. 
\subsection{Toom's interface model}
\label{s:Toom's interface model} Toom's model  is the subject of many theoretical and numerical studies \cite{ToomVasilyevStavskayaMityushinKurdyumovPirogov1990,DerridaLebowitzSpeerSpohn1991,BarabasiAraujoStanley1992,BarkemaFerrariLebowitzSpohn2014,CrawfordKozma2020}. As in the DEP, each site is either empty or occupied, but while in the DEP the jump range is either $1$ or $2$, in Toom's model it is unbounded. 
More precisely, one chooses a spin $i$, and flips it with the next spin to its right which has the opposite sign. 
The rates, for any $k\in \N$, are
\begin{itemize}
\item[(1)$_k$] $\underbrace{1\dots 1}_{k}0 \to 0\underbrace{1\dots 1}_k$ with rate 1,
\item[(2)$_k$]  $\underbrace{0\dots 0}_{k}1 \to 1\underbrace{0\dots 0}_k$ with rate  $r$.
\end{itemize}
When $r=1$, as in the DEP, the particle-hole symmetry is preserved, and we refer to the model as the \emph{symmetric Toom model} (sToom). When $r\neq 1$ this symmetry is broken, and we refer to the model as the \emph{asymmetric Toom model} (aToom).

In Toom's model  with periodic boundary conditions  each stationary state  is   non-interacting, and the total particle current is 
\be
j(\rho)=\rho (1-\rho) + 2 \rho^2 (1 - \rho) + 3 \rho^3(1-\rho)+ \dots - r \left[(1-\rho)\rho + 2(1-\rho)^2 \rho + \dots \right] = \frac{\rho}{1-\rho}-\frac{r(1-\rho)}{\rho}.
\ee
This current vanishes at the critical density $\rho_{\rm c} = \frac{\sqrt r}{1+\sqrt{r}}$.
As thoroughly discussed  in \cite{DerridaLebowitzSpeerSpohn1991}, the system with   closed boundaries and for $L\to \infty$ reaches a stationary state with this density at its left end.

\subsection{Continuum theory}
\label{s:Continuum theory}
As suggested in  \cite{DerridaLebowitzSpeerSpohn1991,BarkemaFerrariLebowitzSpohn2014}, these models can be described using a continuum theory. We define the \emph{height function} $h(x,t)$ as the number of particles   between the left boundary (first spin at site $1$) and $x$, minus its expectation. In the continuum limit we expect it to solve a stochastic differential equation of the type
\begin{equation} \label{eq:general_HL}
\partial_t h(x,t) = c + D \nabla^2 h(x,t) -\mu \nabla h(x,t) + \lambda_2 \big[\nabla h(x,t)\big]^2 + \lambda_3 \big[\nabla h(x,t)\big]^3 + \dots + \sigma \xi(x,t).
\end{equation}
Discarding higher-order terms, we are left with two possible limits: for   sDEP and sToom, the particle-hole symmetry forces the equation to stay invariant under the transformation $h \mapsto -h$, hence the first and fourth terms are not allowed ($c=\lambda_2=0$), and we are left with the \emph{anchored advected Edwards-Wilkinson} equation 
\begin{equation}\tag{aaEW}\label{eq:aaEW}
\begin{split}
\partial_t h(x,t) =&\; D \nabla^2 h(x,t) - \mu \nabla h(x,t) + \sigma \xi(x,t),\\
h(t,0) =&\; 0 , \quad \left<\xi(x,t) \xi(x',t')\right> = \delta(x-x')\delta(t-t').
\end{split}
\end{equation}
In aDEP and aToom the constant and quadratic terms are allowed, leading to the \emph{anchored advected KPZ} equation
\begin{equation}\tag{aaKPZ}\label{eq:aaKPZ}
\begin{split}
\partial_t h(x,t) =&\; D \nabla^2 h(x,t) -\mu \nabla h(x,t) + \lambda_2 \big[\nabla h(x,t)\big]^2 + c + \sigma \xi(x,t),\\
h(t,0) =&\; 0, \quad \left<\xi(x,t) \xi(x',t')\right> = \delta(x-x')\delta(t-t').
\end{split}
\end{equation}
Since $ \left<  h(x,t)\right> =0$ by construction, the first two terms on the r.h.s.\ vanish on average, and hence
\begin{equation} \label{eq:0current}
c = -\lambda_2 \langle [\nabla h]^2 \rangle.
\end{equation}
An additional term with $\lambda_3\neq 0$ may lead to   logarithmic corrections \cite{PaczuskiBarmaMajumdarHwa1992}.

A   direct derivation of this limit for Toom's model appears in \cite{DerridaLebowitzSpeerSpohn1991,BarkemaFerrariLebowitzSpohn2014}. We describe here the derivation for the DEP.

\subsection{Continuum limit of the DEP} \label{sec:DEP_limit}
Let us consider the DEP at the critical density $\rho_c=\frac{r}{1+r}$.
Define $\eta(x,t)$ to be $1$ if there is a particle at $x$ at time $t$ and $0$ otherwise. Then the height function $h$ is defined as $h(x,t):=\sum_{y=1}^x [\eta(y,t) - \rho]$.

During a time period $\dt$, the value of $h$ at $x$   changes when particles jump from the left of $x$ to its right or vice versa:
\begin{enumerate}
\item $h(x,t+\dt) = h(x,t) - 1$ with probability $p_1 = \eta(x,t)[1-\eta(t,x+1)]\dt$,
\item $h(x,t+\dt) = h(x,t) + 1$ w.p.\ $p_2 = [1-\eta(x,t)]\eta(t,x+1)\dt$,
\item $h(x,t+\dt) = h(x,t) - 1$ w.p.\ $p_3 = \eta(x,t)\eta(x+1,t)[1-\eta(x+2,t)]\dt$,
\item $h(x,t+\dt) = h(x,t) - 1$ w.p.\ $p_4 = \eta(x-1,t)\eta(x,t)[1-\eta(x+1,t)]\dt$,
\item $h(x,t+\dt) = h(x,t) + 1$ w.p.\ $p_5 = r[1-\eta(x,t)][1-\eta(x+1,t)]\eta(x+2,t) \dt$,
\item $h(x,t+\dt) = h(x,t) + 1$ w.p.\ $p_6 = r[1-\eta(x-1,t)][1-\eta(x,t)]\eta(x+1,t) \dt$.
\end{enumerate}
First, we calculate  
\begin{equation*}
\langle h(x,t+\dt) - h(x,t) \rangle = (-p_1 + p_2 -p_3- p_4+p_5+p_6 ) \dt.
\end{equation*}
We seperate $\eta(x,t)$ into its average $\rho$ plus fluctuations, $\eta(x,t)=\rho + \nabla h(x,t)$. This yields
\begin{align}
\langle h(x,t+\dt) - h(x,t) \rangle &= \big[ D \nabla^2 h(x,t) - \mu \nabla h(x,t) + \lambda_2 \big(\nabla h(x,t)\big)^2 + c\big] \dt, \\
D &=  1+3r-6r\rho +3r\rho^2 + 3\rho^2, \\
\mu &= -6 r  \rho ^2+8 r  \rho
   -2 r -6 \rho ^2+4 \rho \\
\lambda_2 &= 6 r\rho - 4r + 6\rho - 2. 
\end{align}
The noise term is obtained from 
\begin{equation}
\sigma^2 \dt=
\left\langle
\frac{1}{k} \left[ \sum_{x=1}^k h(x,t+\dt)-h(x,t) \right]^2
\right\rangle.
\end{equation}
The reason we   take a spatial average is to account for correlations:  since each of the transitions (1) and (2) contribute $1$ to the sum and transition (3) which is equivalent to (4), and transition (5) which is equivalent to (6), contribute $2$, we obtain
\begin{equation}
\left\langle
\frac{1}{k} \left( \sum_{x=1}^k h(x,t+\dt)-h(x,t) \right)^2
\right\rangle
= p_1+p_2+4p_3+4p_5. 
\end{equation}
Note that $p_4,p_6$ are   spatial shifts of $p_3,p_5$ hence we must count them only once.
This yields
\be
\sigma^2 = 2 \rho(1-\rho) (1 +2 r +2 \rho-2 r  \rho).
\ee
Combining these two equations, we obtain for $\rho=\rho_c=\frac{r}{1+r}$
\be
D= \frac{4 r +1}{r +1}, \quad \mu = \frac{2 r }{r +1}, \quad \lambda_2 = 2(r-1), \quad \sigma^2= \frac{2 r  (5 r
   +1)}{(r +1)^3}.
\ee
This is model
 \eqref{eq:aaEW} when $r=1$ and \eqref{eq:aaKPZ} when $r\neq 1$.
For the symmetric case this gives
\be\label{14}
D=\frac52, \quad \mu =  1, \quad \lambda_2=0, \quad \sigma^2=\frac32.
\ee

\section{Particle-hole symmetric case -- the aaEW universality class}
\label{Particle-hole symmetric case -- the aaEW universality class}
\label{s:Symmetric case -- the aaEW universality class}
\subsection{Periodic BC versus anchored interface}
As discussed for Toom's model and the DEP, models in the advected Edwards-Wilkinson universality class have a stationary measure which depends   on the boundary. A similar behaviour is observed in the continuum limit. If $h$ is a solution of the diffusion equation \eqref{diffusion-equation} with periodic boundary conditions, then $\tilde{h}(x,t) = h(x-\mu t,t)$ solves the non-advected Edwards-Wilkinson equation, 
\bea
\partial_t \tilde{h}(x,t) &=& \partial_t   h(x-\mu t,t) - \mu  \nabla h(x -\mu t,t) = D \nabla^2 h(x-\mu t,t) + \sigma \xi(x-\mu t,t) \nn\\
&=& D\nabla^2 \tilde{h}(x,t)+\sigma  {\xi}(x,t), \quad \left<\xi(x,t) \xi(x',t')\right> = \delta(x-x')\delta(t-t').
\eea
(We used that $\xi(x,t)$ and $\xi(x-\mu t,t)$ have the same correlations.)
In particular, the stationary state for a periodic  system of size $L$ is a Brownian motion with periodic boundary conditions (i.e., a Brownian bridge). For $L\to \infty$, this change of variables is   valid on the entire line $x\in\R$.
However, if we consider the half line $x \in [0,\infty)$ with Dirichlet boundary condition $h(0)=0$ as in equation \eqref{eq:aaEW}, we cannot define $\tilde{h}$ as above, and  the stationary state is \emph{not} a Brownian. This choice of boundary, corresponding to an anchored interface, is the subject of this article.

\subsection{Bulk behaviour} When considering portions of the interface far away from the origin, the effect of the boundary becomes negligible, and in particular the stationary state on an interval $[x,x+\Delta x]$ looks like a Brownian motion if $\Delta x \ll x$. This is shown rigorously for Toom's model in \cite{CrawfordKozmaRoeck2015,CrawfordKozma2020}, and we expect a similar behaviour in all models of this universality class.

\subsection{Edge behaviour} As mentioned above, unlike the bulk behaviour, the behaviour of the model near the edge is drastically different from the standard Edwards-Wilkinson universality class \cite{DerridaLebowitzSpeerSpohn1991,BarkemaFerrariLebowitzSpohn2014,Pruessner2004}. The combination of the Dirichlet boundary and the advection term transfers temporal correlations into spacial ones. In the following, we derive analytic results for the edge behaviour in the anchored case.

\subsection{Basic formulas}
We solve the advected diffusion equation with Dirichlet boundary conditions, see also \cite{Pruessner2004}\footnote{In \cite{Pruessner2004} $\mu$ has opposite sign, which is equivalent to studying the equation on the negative half line $(-\infty,0]$.}. By simple rescaling we write equation \eqref{eq:aaEW} in the form:
\begin{equation}\label{e1}
\begin{split}
&\partial_t h(x,t) = D\nabla^2 h(x,t) - \mu \nabla h(x,t) + \xi (x,t),\qquad h(x=0,t) = 0.\\
& \left< \xi (x,t) \xi(x',t')\right> = \delta(x-x') \delta(t-t').  
\end{split}
\end{equation}
\Eq{e1} for $x>0$ is solved by
\be\label{4}
h(x,t) = \int_{y>0} \int_{t'<t} P_{\mu}(x,t|y,t') \xi(y,t'). 
\end{equation}
Here
\bea
\label{prop}
P_0(x,t) &=& \frac{e^{-\frac{x^2}{4D t}}}{\sqrt{4 D\pi t}}, \nn \\
P^\mu_0(x,t) &=& P_0(x,t) \,\rme^{\frac {\mu x}{2D}-\frac{\mu^2 
t}{4D}},\\
P^\mu(x,t|y,t') &=& \big[ P_0  (|x-y|,t-t')-P_0 (x+y,t-t')\big]  \,\rme^{\frac {\mu (x-y)}{2D}-\frac{\mu^2 (t-t')}{4D}}. 
\nn
\eea

\subsection{Analytic result for the variance of the height}
\label{s:Analytic result for the variance of the height}
The roughness exponent $\zeta$ is given by the behaviour of $\left< h(x,t)^2\right>$ in the steady state, 
\bea
\left< h(x,t)^2\right>  &=& \int_{y_1>0} \int_{t_1<t} P^{\mu}(x,t|y_1,t_1)   \int_{y_2>0} \int_{t_2<t} P^{\mu}(x,t|y_2,t_2)  \left<\xi(y_1,t_1) \xi(y_2,t_2)\right> \nn\\
&=& \int_{y>0} \int_{t'<t} P^{\mu}(x,t|y,t')^2 . \label{advEWana1}
\eea
This is independent of $t$, and we will drop $t$ from now on. 
To proceed, we use the Fourier-transform of the propagator \eq{prop}, 
\bea
P^{\mu}(x,t|y,0) &=&
 \frac{\rme^{-\frac{\mu ^2 t}{4D}+\frac{\mu  (x-y)}{2D} }}{ 
   \sqrt{4 D \pi  t}} \left[\rme^{- \frac{(x-y)^2}{4 Dt}} -\rme^{-\frac{(x+y)^2}{4 Dt}} \right] \nn\\
&=&\int \frac{\rm d k}{2\pi} \left[\rme^{i k (x-y)}-\rme^{i k
   (x+y)}\right] \rme^{ -k^2 D t-\frac{\mu ^2
   t}{4D}+\frac{  \mu }{2D}
   (x-y)}  . \qquad 
\eea
From now on, we set $\mu=D=1$. 
We can recover the full $\mu$, $D$ and $\sigma$-dependence by remarking that 
\be\label{mu-dep}
\left< h(x)^2\right>_\mu = \frac{\sigma^2}\mu  \left< h\Big(\frac{\mu x}{D},\frac{\mu^2 t}D\Big)^2\right>_{\mu=D=\sigma=1}.
\ee
Then
\be\label{62}
\int_{y>0}\int_{t>0} P^{\mu=1}(x,t|y,0)^2   
= \int_{k,p}\frac{-16   \, k   p \rme^{x [1
   +i (k+p)]}}{ [2(k^2+p^2 )+1]  [(k-p)^2+1 ]  [(k+p)^2+1 ] }.
\ee
Integrating \Eq{62} over $k$ yields 
\be\label{61}
\left< h(x)^2\right>_{\mu=1} = \int_{-\infty}^{\rm \infty}\frac {\rmd p}{2\pi i} \frac{16  p \rme^{-\sqrt{p^2+\frac{1}{2}}
   x+i p x+x}}{ (4 p^2+1 )^2} .
\ee
\Eq{61} can be simplified, using for the denominator
\be
\int_0^\infty \rmd t\, t \,\rme^{-a t} = \frac1{ a^2},
\ee
and for $\rme^{-\sqrt{p^2+\frac{1}{2}} x}$
\be
\int_{s>0}
\frac{\rme^{-a s-\frac{1}{4 s}}}{2
   \sqrt{\pi } s^{3/2}} =\rme^{-\sqrt a}.
\ee
Then the integral over $p$ can be done, leading to 
\be
\left< h(x)^2\right>_{\mu=1}  = \int_{s>0}\int_{t>0}  \frac{2 t x \,\rme^{-\frac{x^2}{4 s x^2+16
   t}-\frac{s x^2}{2}-\frac{1}{4
   s}-t+x}}{\pi  s^{3/2}  (s x^2+4
   t )^{3/2}}.
\ee
Setting $s \to 4st/x^2 $, and integrating over $t$ yields 
\be
\left< h(x)^2\right>_{\mu=1}  = \int_{s>0}\frac{\rme^x x^2 K_0\left(\frac{(2 s+1) x}{2
   \sqrt{s} \sqrt{s+1}}\right)}{4 \pi 
   s^{3/2} (s+1)^{3/2}} .
\ee
$K_0$ is the Bessel $K_0$-function.
This can further be simplified, by first setting $y=\sqrt{  s/({1+s})}$, and second $2u = y+y^{-1}$.
The final result is 
\be
\left< h(x)^2\right>_{\mu=1}  =\frac {x^2 \rme^{x}}{\pi} \int_1^{\infty}  K_0(x u)\, {\rmd u} = \frac {x \rme^{x}}{\pi} \int_x^{\infty}  K_0(u)\, {\rmd u} .
\ee
The integral is known analytically, 
\be
\left< h(x)^2\right>_{\mu=1}  =  \frac{x \,\rme^x }{2}  \Big[ 1-x \Big(\pmb{L}_{-1}(x)
   K_0(x)+\pmb{L}_0(x) K_1(x) \Big) \Big].
\ee
The function $\pmb{L}_n(x)$ is the modified Struve-function. 
For $\mu\neq1$ this reads according to \Eq{mu-dep}
\be
 { \left< h(x)^2\right>  =\frac {x \,\rme^{\mu x}}{\pi} \int_{\mu x}^{\infty}  K_0(u)\, {\rmd u} } .
\ee
One can   approximate the Bessel function for large argument as 
\be
K_0(x) = \rme^{-x}\left[ \sqrt{\frac{\pi }{2 x}}+\ca O(x^{-\frac 32}) \right].
\ee
Then the integration can be done analytically, 
\be\label{23}
\left< h(x)^2\right>  \simeq \frac{e^{\mu x} x\,
   \text{erfc}\left(\sqrt{\mu x}\right)}{\sqrt
   {2}} = \frac{\sqrt{x}}{\sqrt{2 \pi\mu }} + \ca O(x^{-\frac12}).
\ee
This gives a roughness exponent 
\be
\zeta= \frac14.
\ee

\subsection{Approximate calculation for large $x$}
\begin{figure}
\centerline{\fig{0.6}{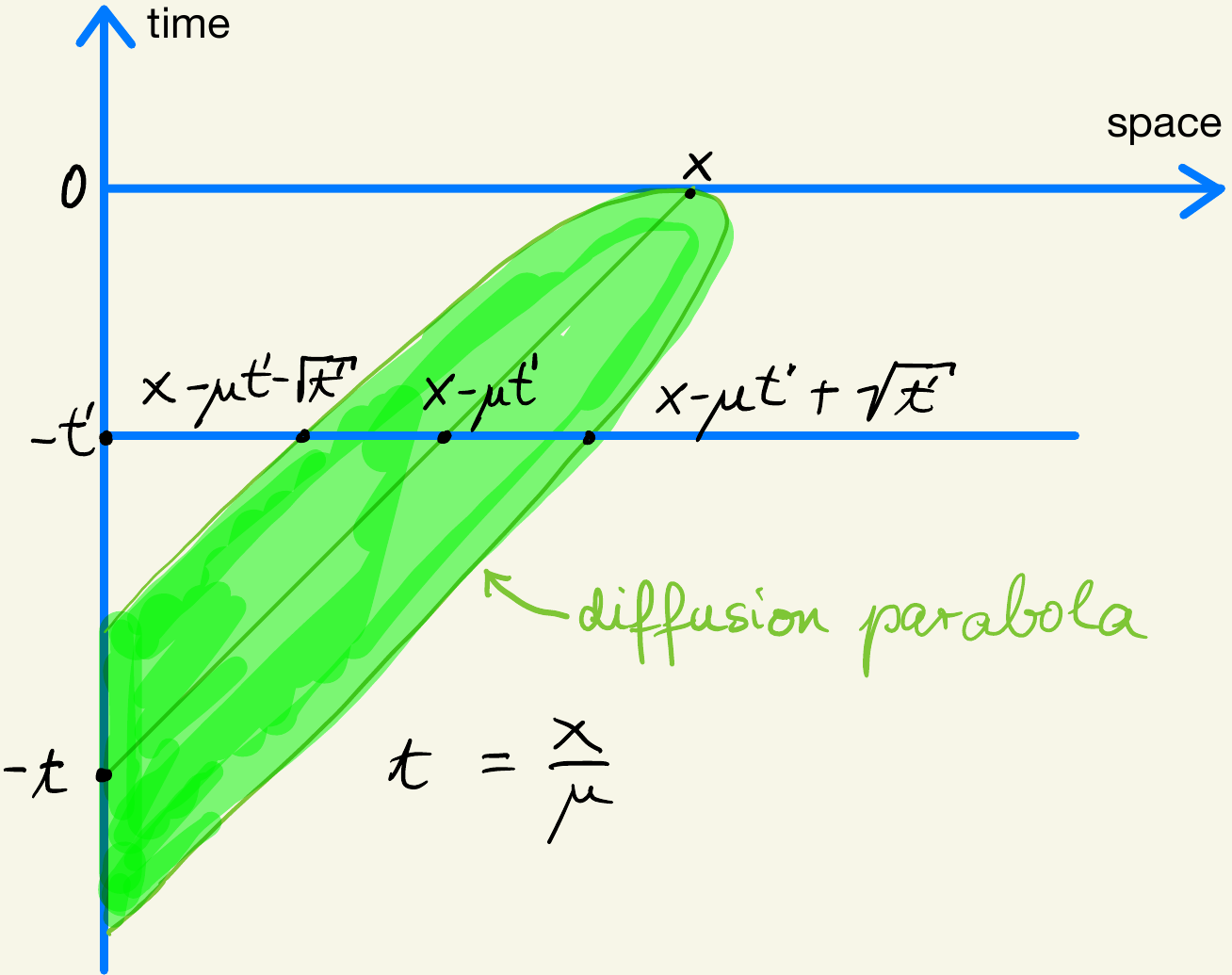}}
\caption{Sketch of the domain contributing to the integral \eq{P-approx}. Support in space at given time $t'$ is restricted to the {\em diffusion parabola} as indicated. This implies that for $0<t'<t$ the Dirichlet propagator can be approximated by the free propagator,  and the  space integral   extended to $\infty$. By the same argument, times $t'>t$ do not contribute.}
\label{fig:diffusion_parabola}
\end{figure}
From Figure \ref{fig:diffusion_parabola} we see that, for large $x$, we can approximate \Eq{4} as 
\bea\label{P-approx}
h(x,t=0) &=& \int_0^\infty \rmd y \int_{-\infty}^{0}\rmd t' P^{\mu}(x,t|y,t') \xi(y,t') \nn\\
&\approx& \int_{-\infty}^\infty \rmd y \int^0_{-\frac x\mu } \rmd t'\,P_0^\mu(x-y,-t') \xi(y,t')\nn\\
&=& \int_{-\infty}^\infty \rmd y \int_0^{\frac x\mu } \rmd t'\,P_0^\mu(x-y,-t') \xi(y,t'). 
\eea
Then 
\bea\label{26}
\left< h(x)^2\right> &\approx&  \int_{-\infty}^\infty \rmd y \int_0^{\frac x\mu } \rmd t'\,P_0^\mu(x-y,t')^2 \nn\\
 &=&  \frac{1}{ \sqrt{8\pi}} \int_0^{\frac x\mu }   \frac{\rmd t'}{\sqrt{t'}} = \sqrt{\frac{x}{2\pi \mu }}.
\eea
This reproduces the result of \Eq{23}.

\subsection{Decay of the auto-correlation function}
\label{s:Decay of the auto-correlation function}
We wish to calculate (with $\mu=D=1$)
\be
\left< h(x,t) h(x,0) \right>   =  \int_{y>0} \int_{t'<0} P_{\mu}(x,0|y,t') P_{\mu}(x,t|y,t') . \label{advEWana2}
\ee
With the help of \Eq{P-approx}, this can be approximated as 
\bea
\left< h(x,t) h(x,0) \right>  
&\approx&  \int_y \int_{0\le t'\le x} P_{0}^\mu(y,t') P_{0}^\mu(y+t,t+t').
\eea
Therefore 
\bea
\!\left< h(x,t) h(x,0) \right>  &\simeq &\int_{-\infty}^\infty \frac{\rmd k}{2\pi}\int_{-\infty}^\infty \frac{\rmd p}{2\pi} \int_y \int_{0\le t'\le x} \rme^{ik(x-y+t')  -k^2t'} \rme^{i p (x-y+t+t') -p^2 (t+t')}\nn\\
&& = \int_{0\le t'\le x} \int_{-\infty}^\infty \frac{\rmd k}{2\pi} \rme^{ -k^2 (t+2t') -i k t } =\int_{0\le t'\le x} \frac{\rme^{-\frac{t^2}{4(t+2t')}}}{2 \sqrt{\pi} \sqrt{t+2 t'}} \nn\\
&&=\frac1 4 \sqrt{\frac t  \pi}\int_{0}^{\frac {2 x }t}\rmd y\, \frac{\rme^{-\frac{t}{4(1+y)}}}{ \sqrt{1+y}} = \frac t{8 \sqrt{\pi} }\left[ \Gamma\Big(-\frac12,\frac{t^2}{4(t+2x)} \Big) -\Gamma\Big(-\frac12,\frac{t}{4} \Big)\right] . \qquad 
\eea
The $\Gamma$-function decays as a power-law corrected Gaussian for large second argument. As we are interested in both large $x$ and large $t$, the second term vanishes; the first one implies that the proper scaling variable is $t/\sqrt x$, reducing the denominator of the first term from  $t+2x\to 2x$. As a result, 
\be\label{f-final}
\begin{split}
\left< h(x,t) h(x,0) \right>   &\simeq \sqrt{\frac{{x}}{ {2\pi}}}
\, f\!\left(\frac t{\sqrt x}\right),
\quad f(0)=1, \quad f(y) \stackrel{y\to \infty}{-\!\!\!\longrightarrow} 0,  \\
f (y) 
&= \frac{y  
   }{4 \sqrt{2}} \,\Gamma
   \Big(-\frac{1}{2},\frac{y^2}{8}\Big).
\end{split}
\ee
We tested  this against a numerical integration of the exact expression. This worked, but only for large $x$ and $t$; for small times our approximation converges from the wrong side.
$f(y)$ has series expansions for small and large $y$
\be
\begin{split}
f(y) &= 1-\frac{1}{2} \sqrt{\frac{\pi }{2}}
   y+\frac{y^2}{8}-\frac{y^4}{384}+\frac{y
   ^6}{15360}
   + ... \\
f(y)   &=  \rme^{-\frac{y^2}8}\left[\frac{4}{y^2}-\frac{48}{y^4}+\frac{960}{y^6}+ ... \right] .
\end{split}
\ee
Putting back the dependence on  $\mu$,  $D$ and $\sigma$  yields with \Eq{mu-dep}  
\be \label{eq:phi_correlation}
\left< h(x,t) h(x,0) \right> \simeq \sigma^2 \sqrt{\frac{x}{2\pi \mu D}}\, f\!\left(\frac{\mu^2 t}{\sqrt{D x \mu}}\right).
\ee

\subsection{Dynamic exponents $z=2$, $z=1$ and $z=1/2$}
Our calculations above teach us that 
\be
z= \left \{ 
\begin{array}{cl}
2& \mbox{ for bulk observables in the comoving (advected) frame} \\
1& \mbox{ for bulk observables in the fixed frame} \\
\frac12 & \mbox{ for } \left< h(x,t) h(x,0) \right>
\end{array}
\right.
\ee
In particular, we stress that at a time scale $t\sim L$ we reach stationarity on the interval $[0,L]$ (see also \cite{CrawfordKozmaRoeck2015,CrawfordKozma2020}).


\subsection{Numerical verification}
In the symmetric case, we expect the (centered) height function $h$ in both Toom's model and the DEP to converge to  the solution of the aaEW equation \eq{diffusion-equation} (up to possible logarithmic corrections). In order to test this, we compare the correlation $\langle h(x,t)h(x,0) \rangle$ to Eq.~\eqref{eq:phi_correlation}.

\begin{figure}[t]
\fig{0.48}{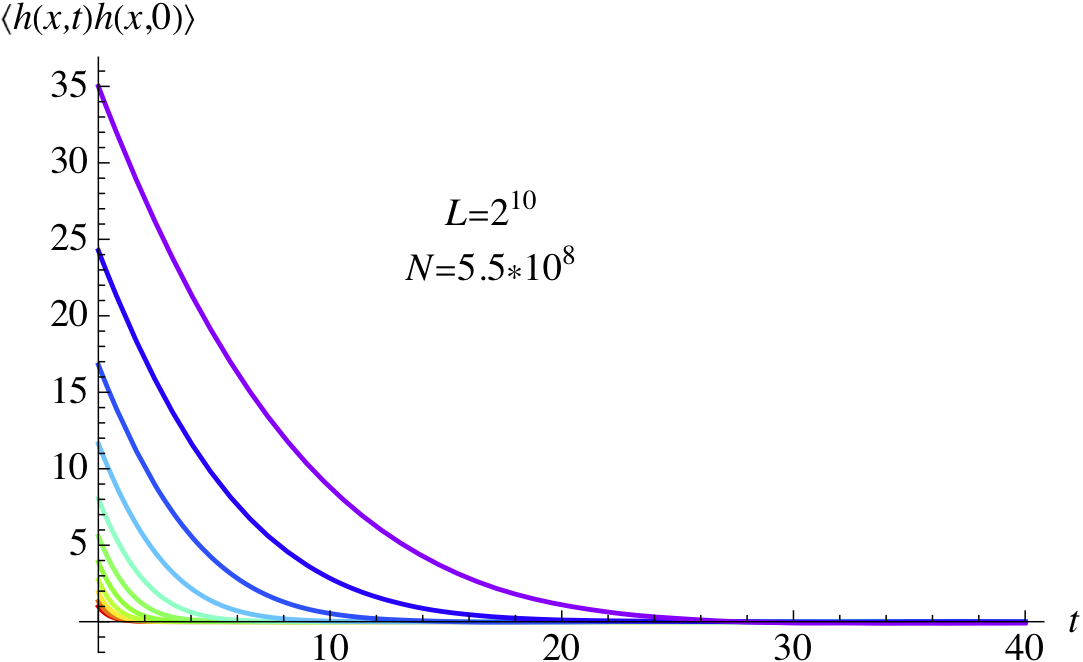}\hfill\fig{0.48}{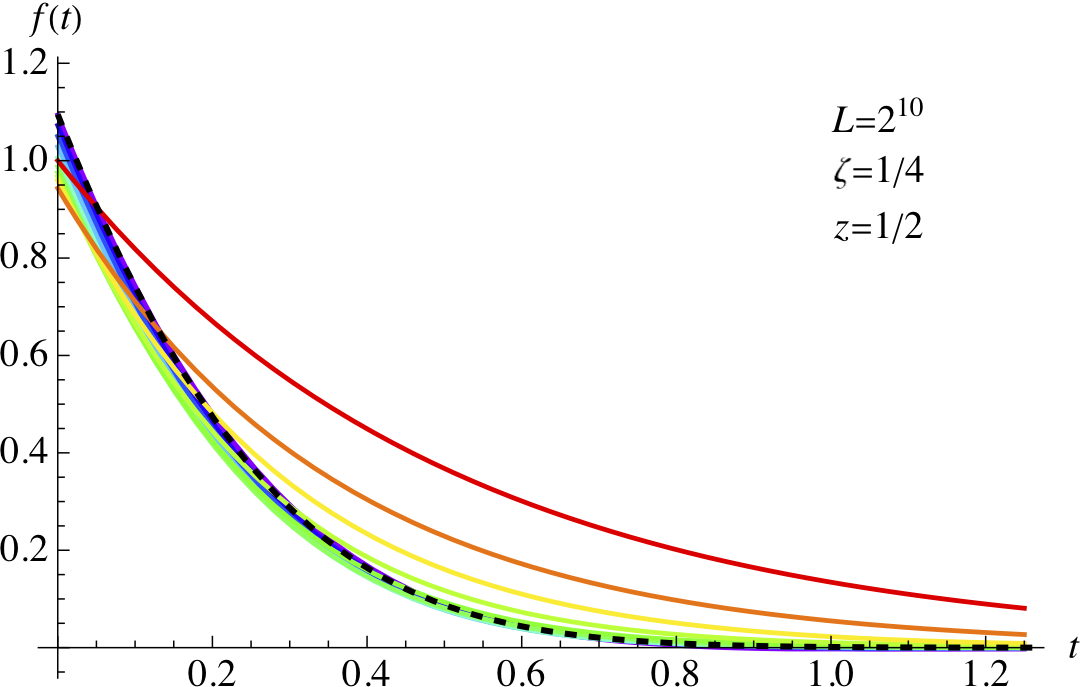}\hfill
\fig{0.48}{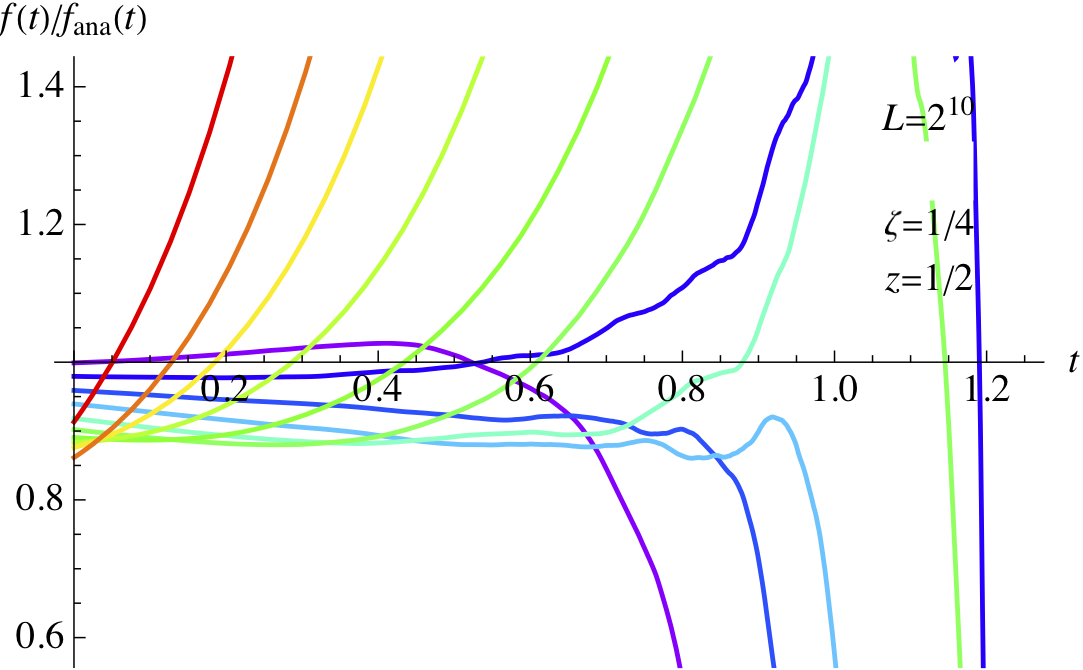}\hfill
\fig{0.48}{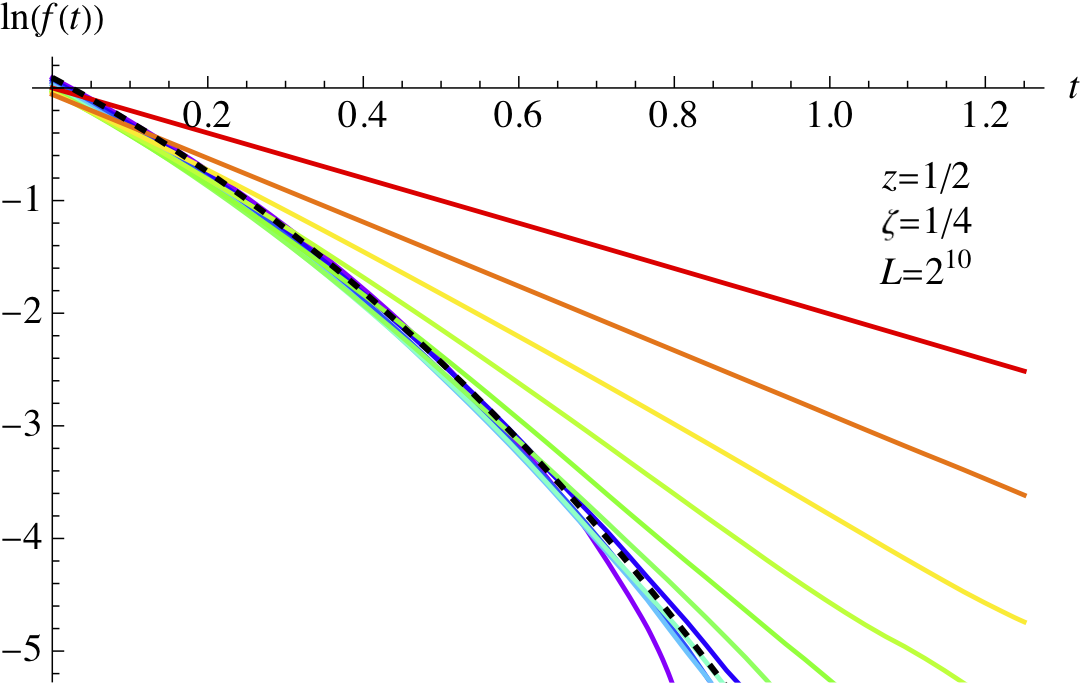}\hfill
\caption{Simulation of the height correlations in the symmetric Toom model of size $L=2^{10}$, after $N=5.5\cdot 10^7$ iterations. Top left: Height-correlation function for $x=2$ (small, red), $x=4$, $x=8$, ...  to $x=2^{10}$ (largest, violet). 
Trop right: convergence (same color code) of the scaling function $f(t)$ against the function of \Eq{f-final} (black dashed), with two arbitrary scales. Bottom left: the ratio of measured $f(t)$ to analytic prediction.  Bottom right: $f(t)$ on a log-scale.}
 \label{fig:toom_correlation}
\end{figure}

In Fig.~\ref{fig:toom_correlation} we show this comparison for Toom's model. It is worth noting, as already mentioned in \cite{DerridaLebowitzSpeerSpohn1991}, that there are no finite-size effects in this simulation. 
We see that the analytic prediction \eq{eq:phi_correlation} is well verified. (We did not check the dependence on  parameters $\mu$, $D$, and $\sigma$).
For more thorough numerical studies of this model we refer the reader to \cite{DerridaLebowitzSpeerSpohn1991,BarkemaFerrariLebowitzSpohn2014,SubramanianBarkemaLebowitzSpeer1996}.)

\begin{figure}[bt] 
\fig{0.48}{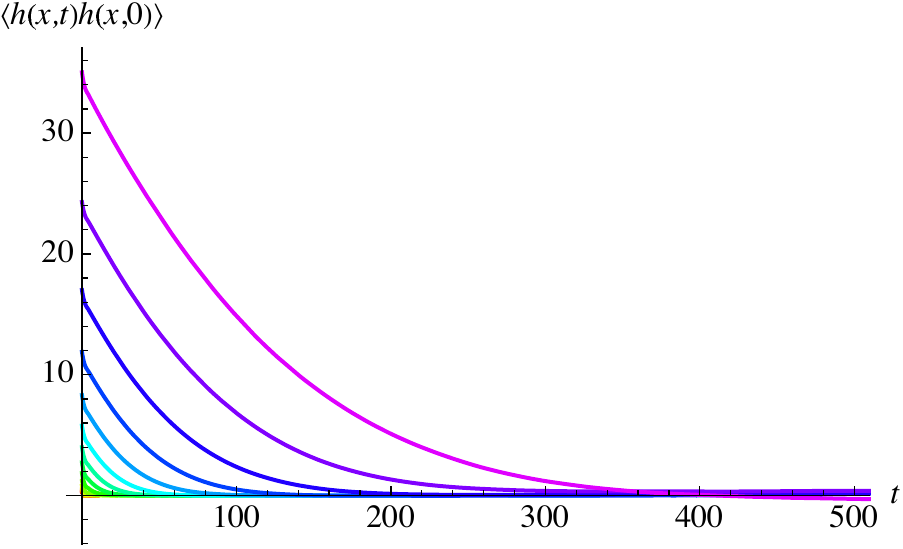}\hfill
\fig{0.48}{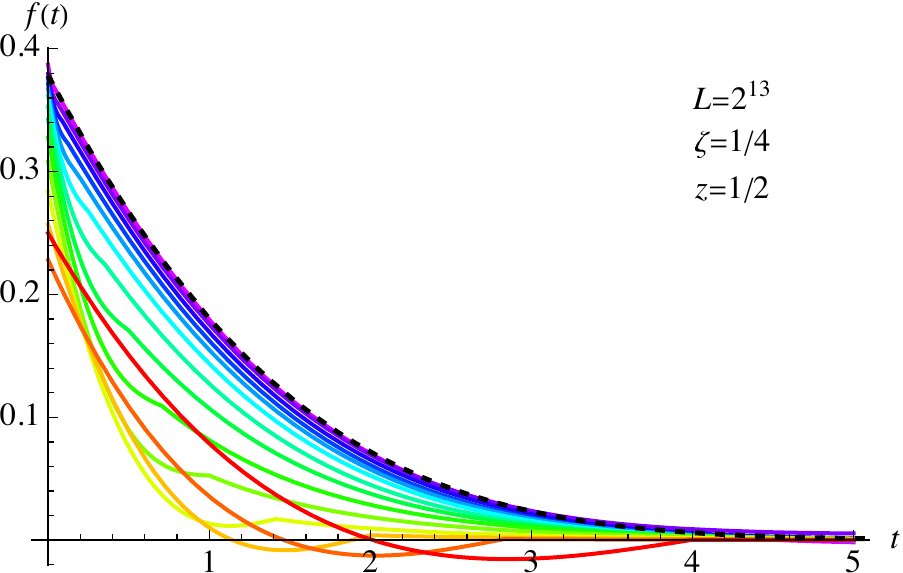}\hfill
\fig{0.48}{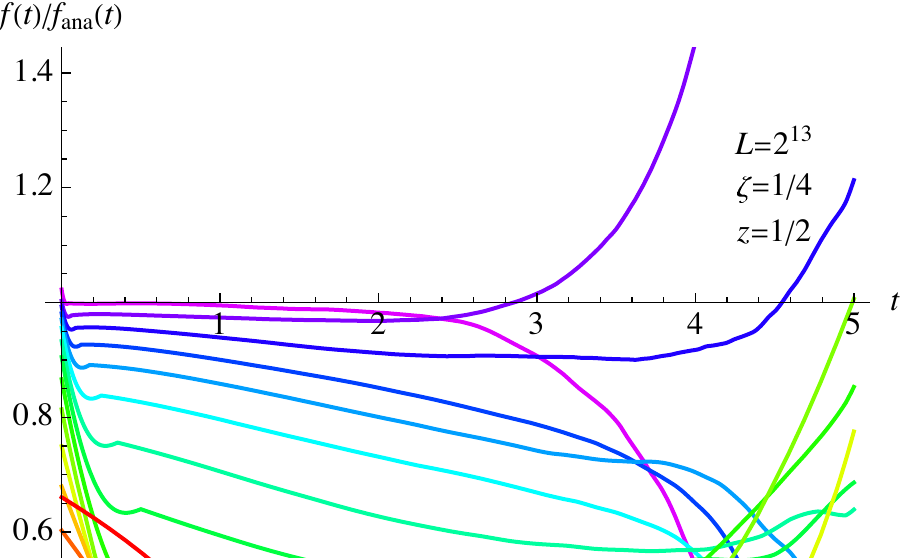}\hfill
\fig{0.48}{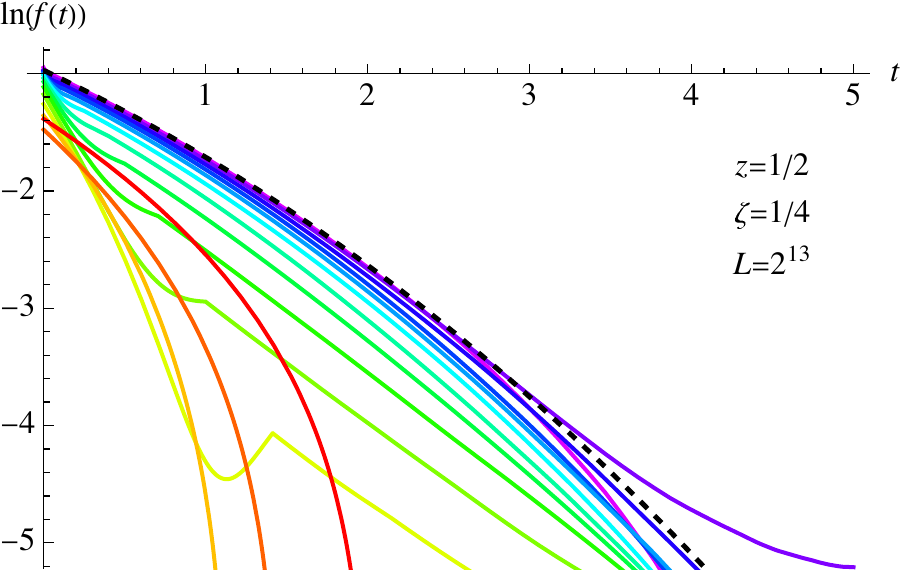}\hfill
\caption{Simulation of   sDEP for $L=2^{13}$, after $N=5 \cdot 10^{12}$ iterations.  Top left: Height-correlation function for $x=2$ (small, red), $x=4$, $x=8$, ...  to $x=2^{13}$ (largest, violet). Trop right: convergence (same color code) of the scaling function $f(t)$ against the function of \Eq{f-final} (black dashed), with two fitted scales, and dropping the first data point. The amplitude is as predicted in \Eq{48}. Bottom left: the ratio of measured $f(t)$ to analytic prediction.  Bottom right: $f(t)$ on a log-scale.}
\label{fig:sdep_correlation}
\end{figure}

In Fig.~\ref{fig:sdep_correlation} we show the same plot for the symmetric DEP. There are noticeable finite-size effects, but for large systems the rescaled correlator $\left< h(x,t) h(0,0\right> x^{-1/2}$ plotted against $t/x^2$ seemingly converges.
Remarkably, we obtain not only the correct exponent $\zeta = \frac{1}{4}$, but also the correct coefficient: after rescaling Eq.~\eqref{e1} to account for the coefficients $D$ and $\sigma$ in Eq.~\eqref{eq:aaEW}, and plugging in $D,\mu$ and $\sigma$ found in Section \ref{sec:DEP_limit}, we obtain
\[
\langle h(x)^2 \rangle \approx \frac{\sigma^2}{\sqrt {2\pi    \mu    D}}\sqrt x = \sqrt{\frac{9}{20 \pi}} \sqrt x \approx 0.378 \sqrt x.
\] 
This relation, including the coefficient of $0.378$   is verified numerically, as can be seen on Fig.~\ref{fig:sdep_correlation}. 
On the other hand, the rescaling factor for the argument of $f$ is off by a factor of about $1.66$, 
\be\label{48}
\left< h(x,t) h(0,0)\right>  x^{-1/2} \approx  \sqrt{\frac{9}{20 \pi}}  f \left( {1.66} \sqrt {\frac25}\frac t{\sqrt{x}} \right).
\ee
This factor, which seemingly  affects time scales only,  may be due to the presence of sub-sub-leading corrections $\sim (\nabla h)^3$ in the equation of motion.
The presence of this type of corrections has been discussed in Refs.~\cite{PaczuskiBarmaMajumdarHwa1992,SubramanianBarkemaLebowitzSpeer1996}. As a marginal perturbation $ (\nabla h)^3$  
gives logarithmic corrections, which may be of the form $\ln L$ or $\ln t$.
Our simulation suggests that the combination $\frac{\sigma^2}{\sqrt{\mu D}}$ does not renormalize, while each coefficient by itself may renormalize. 
In particular the combination $\mu^3/D$ renormalizes by a factor of $2.8$ at $L=2^{13}$. We show in section \ref{s:Protection of the stationary measure} that $\sigma^2/D$ does not renormalize 
in the bulk. In summary, static observables seem not to renormalize, while temporal correlations do.

The code used to simulate the DEP is available as a part of the arXiv version of this paper.

\subsection{Discrete argument}
The exponent $z=\frac{1}{2}$  in the aaEW universality class can also be derived directly in the discrete model, without passing to the continuum limit.

Consider the height function  $h(L)$ of the sDEP at   fixed position $L$. This function   only changes when a particle or a hole at $L$ or $L+1$ jumps to the right of $L$. 
Since the density is approximately $1/2$, at any such jump $h(L)$ increases by $1$ with probability close to $1/2$ and decreases by $1$ with probability close to $1/2$. However, due to density fluctuations, this probability is not exactly $1/2$. When the density is slightly above $1/2$ there are more particles, and $h$ is more likely to decrease. Conversely, when the density fluctuates below $1/2$, it is more likely to increase. The exact correction is     complicated, depending on non-trivial correlations, but to first order we may assume that it induces a drift $-\alpha  h(L)$ with $\alpha>0$, possibly $L$-dependent.
We conclude that $h(L)$ behaves as a random walk   with a drift term $-\alpha h(L)$, equivalent to diffusion in a confining potential $\frac{\alpha}{2} h(L)^2$. For this process, we know that $h(L)$ fluctuates on a scale $\alpha^{-1/2}$ and that its relaxation time is of   order $1/\alpha$.
Finally,   since $\zeta=\frac{1}{4}$, the fluctuations of $h$ are of   order $L^{1/4}$. Comparing this scaling with $\alpha^{-1/2}$ suggests that $\alpha$ scale as $L^{-1/2}$, and hence $z=\frac{1}{2}$.

\section{Asymmetric case -- the aaKPZ universality class}
\label{s:Asymmetric case -- the aaKPZ universality class}
\subsection{Introduction}
\label{s:intro:asym}
The aaKPZ universality class can be analysed in a similar manner.
As in the symmetric case, the stationary state in the periodic or infinite system is that  of   Brownian motion, and the bulk behaviour is determined by the KPZ equation in the comoving frame.
The edge behaviour in the aaKPZ universality class can then be studied using methods close to those described above \cite{DerridaLebowitzSpeerSpohn1991,KrugSocolarGrinstein1992,KrugSocolar1992,Krug1997}, yielding a roughness exponent $\zeta = \frac{1}{3}$, and dynamical exponents $z=\frac{3}{2}$ for bulk observables in the comoving frame, $z=1$ for bulk observables in the fixed frame, and $z=\frac{2}{3}$ for $\langle h(x,t)h(x,0)\rangle$.

\subsection{Protection of the stationary measure}
\label{s:Protection of the stationary measure} 
Consider the equation of motion \eq{e1}, with a subleading KPZ-term, and a subsubleading term $\sim (\nabla h)^3$, \footnote{Here we extend an argument initially given for the KPZ equation.  While it may be well known in this more general form, we could not find a reference in the literature. J.~Krug mentions it in his Diplomarbeit of 1985 at Munich university.}
\be\label{49}
\eta \partial_t h(x,t) = D\nabla^2 h(x,t) - \mu \nabla h(x,t) + \lambda_2 [\nabla h(x,t)]^2 + \lambda_3  [\nabla h(x,t)]^3 +...+ \xi (x,t).
\ee
In the absence of boundaries, the measure $\ca P_t[h]$ satisfies the Fokker-Planck equation  
\be\label{P-ss}
\eta\partial_t \ca P_t[h] = \sigma^2 \int_x \frac{\delta^2}{\delta h(x)^2} \ca P_t[h] - \int_x \frac{\delta}{\delta h(x)}
\left\{ \left(D \nabla^2 h(x) - \mu \nabla h(x,t)+ \sum_{n \ge 2}\lambda _{n} \left[ \nabla h(x) \right]^n \right) \ca P_t[h] \right\}.
\ee
Similar to what happens for KPZ, (see e.g.~\cite{Wiese2021}, section 7.12), 
for $\mu = \lambda_i=0$, a steady-state solution $\partial_t \ca P^{\rm ss}[h] =0$ can be found by asking that 
\be
  \sigma^2  \frac{\delta}{\delta h(x)} \ca P^{\rm ss}[h] =   D \nabla^2 h(x)  \ca P^{\rm ss}[h]  .
\ee
This is solved by 
\be
\ca P^{\rm ss}[h] =  \ca N \exp\left({-\frac{\sigma^2}{2D} \int_x [\nabla h(x)]^2} \right).
\ee
What is the effect of the additional terms?
Inserting the steady state into \Eq{P-ss} yields
\bea
\eta\partial_t \ca P^{\rm ss}[h] &=& - \int_x \frac{\delta}{\delta h(x)}
\left\{ \left( - \mu \nabla h(x,t)+ \sum_{n\ge 2} \lambda _n \left[ \nabla h(x) \right]^n \right) \ca P^{\rm ss}[h] \right\} \nn\\
&=& \frac{\sigma^2}{D } \int_x \nabla^2 h(x) \left( - \mu \nabla h(x,t)+  \sum_{n \ge 2}\lambda _{n} \left[ \nabla h(x) \right]^n\right) \ca P^{\rm ss}[h]  \nn\\
&=& \frac{\sigma^2}{D } \int_x \nabla  \left( - \frac\mu2  [\nabla h(x,t)]^2+  \sum_{n \ge 2} \frac{\lambda _n}{n+1} \left[ \nabla h(x) \right]^n \right) \ca P^{\rm ss}[h] =0. \qquad 
\eea
Note that by going from the first to the second line, we have dropped the   derivative of the terms inside the big round brackets, since they are a total derivative.  The last line vanishes, again due to the fact that this is a total derivative.

This calculation shows that as long as we consider a system without boundaries, the steady state is independent of $\mu$  and   $\lambda_n$. 
When we consider the system from an RG perspective, this means   that $\sigma^2/D$ is not renormalized. There is nothing, however, to protect  $\eta/D$. 
In the presence of a boundary, we do not expect bulk properties to renormalize differently, thus the effective $\eta$, $D$, $\mu$, and $\lambda_i$ to be unchanged if a boundary is introduced.

Let us  mention, that from an RG perspective there may  appear additional renormalizations  on the boundary, and since $\left< h(x,t) h(x,0)\right>$ depends on the distance $x$ to the boundary, they could appear there. Since   the static 2-point function $\left< h(x,t)^2 \right>$ shows no corrections, and temporal-derivative terms are marginal in the bulk, we do not expect this to be the case.   It would be interesting to render this intuitive argument more rigorous.

\section{Oslo model}\nopagebreak
\label{s:Oslo model}\nopagebreak
\subsection{Definition}\nopagebreak
\begin{figure}
\begin{center}
\fig{0.33}{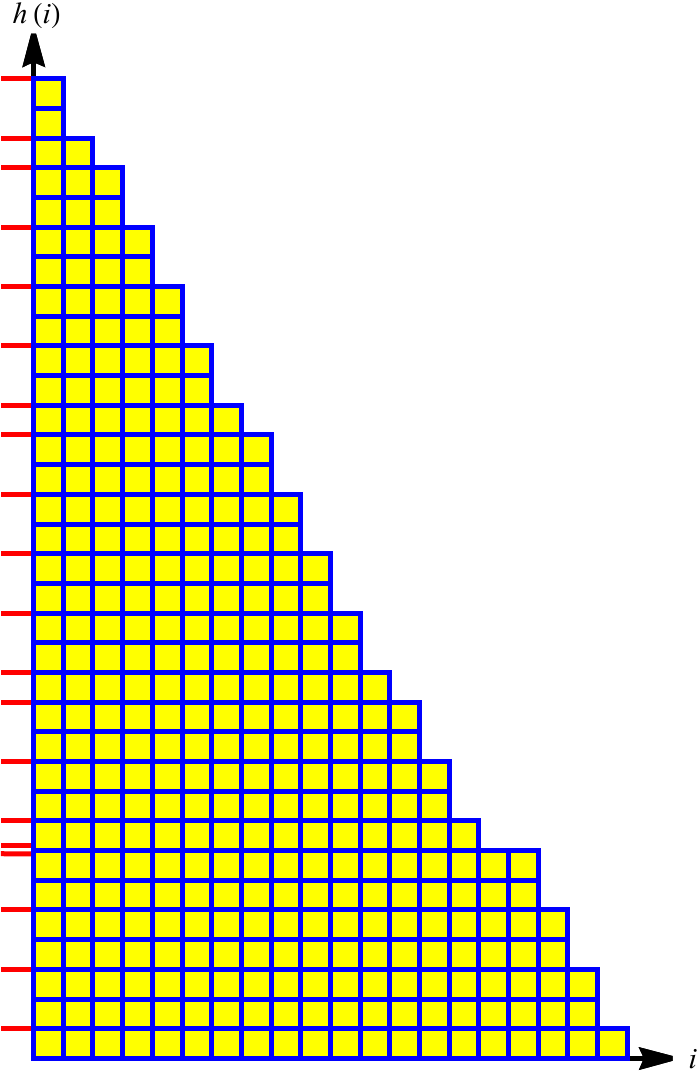}
\parbox{0mm}{\hspace*{0\textwidth}{\raisebox{20mm}[0mm][0mm]{~\;{\fig{0.55}{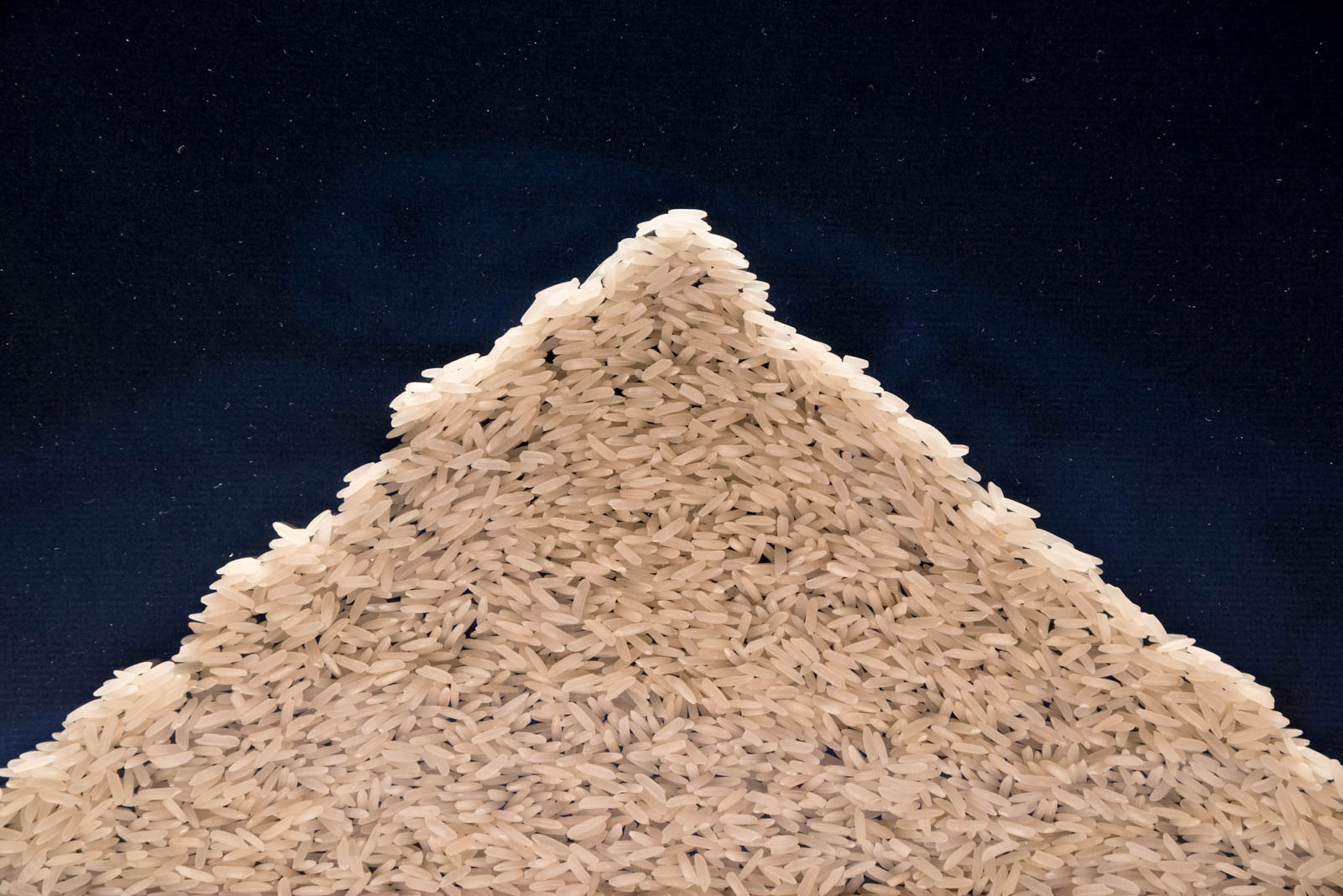}}}}}\fig{0.33}{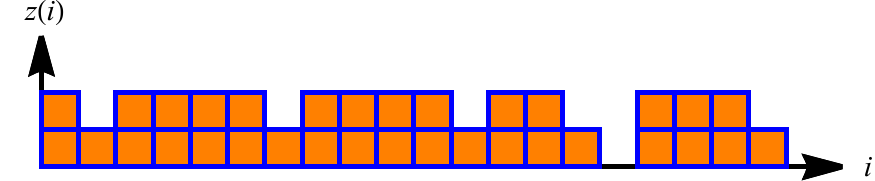}\fig{0.33}{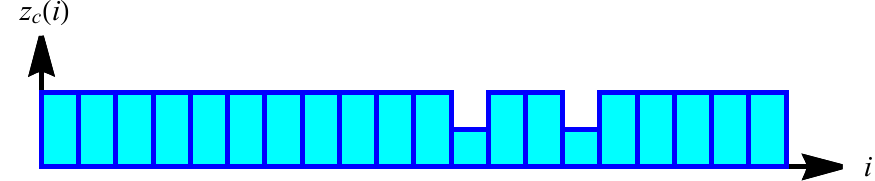}
\end{center}
\caption{Left: a stable configuration of the Oslo model. The latter is a cellular automaton version of the right half of the rice pile at the top. Middle bottom: the variable $z$ defined in \Eq{z-def}. Right: the critical value of $z$ for the stable configuration in the middle.}
\label{f:Oslo-model}
\end{figure} 
The Oslo model describes the evolution of a sand pile, given by its height $h$ (number of grains) at each horizontal position (see Fig.~\ref{f:Oslo-model}). In a real sandpile, whether a grain  at site $i$      slides downhill  depends on the local slope $z(i)$, the friction  with its neighbors of contact, and its orientation. 
The Oslo model is a simple 1-dimensional model for this phenomenon, which depends only on the local slope $z(i)$, and a random variable $z_{\rm c}(i)$. 
It was introduced in 
\cite{Frette1993,ChristensenCorralFretteFederJossang1996}, and is  defined as follows:
Consider the height function $h(i)$ in the left of Fig.~\ref{f:Oslo-model}.
To each height profile $h(i)$ associate a stress field (slope) $z(i)$ defined by 
\be\label{z-def}
z(i) := h(i)-h(i+1).
\ee
In addition, at each position there is a   threshold $z_{\rm c}(i)$.
A toppling 
is invoked if $z(i)>z_{\rm c}(i)$, $i>1$.
The toppling rules are  
\bea
z(i) \to z(i)-2 , \quad
z(i\pm 1)\to z(i\pm 1)+1 . 
\eea
They can be interpreted as moving a grain from the top of the column at site $i$ to the top of the column at site $i+1$, 
\be
h(i)\to h(i)-1,\quad h(i+1)\to h(i+1)+1.
\ee
After such a move, the threshold $z_{\rm c}(i)$ for site $i$  is updated,
\bea
 z_{\rm c}(i) \to \mbox{new random number} .
\eea
In its original version, the random number is $1$ or $2$ with probability $1/2$. To obtain Fig.~\ref{f:Oslo-model} we used a random number drawn uniformly from the interval $[0,2]$. This reduces the critical slope.

In this article we consider a sysmtem of size $L$ with absorbing (Dirichlet) boundary conditions on the left boundary (site $1$) and free (Neumann) boundary conditions on the right boundary (site $L$).
That is, in the $h$ variable, a toppling at $1$ happens when $z(1)>z_{\rm c}(1)$, and it moves a grain from site $1$ to site $2$. On the right boundary, a toppling at $L$ happens when $h(L)>z_{\rm c}(L)$, and it removes one grain from $h$.
We can formulate this in the language of the $z$ variables: when a toppling at $1$ occurs, $z(1)$ decreases by $2$ and $z(2)$ increases by $1$ (i.e., the particle that should have gone to the left exits the system). When a toppling at $L$ occurs, $z(L)$ decreases by $1$ and $z(L-1)$ increases by $1$ (i.e., the particle that should have gone to the right stays at $L$).

For further reading on the Oslo model, we refer to Refs.~\cite{Dhar2004,HuynhPruessnerChew2011,GrassbergerDharMohanty2016,Hinrichsen2000,HenkelHinrichsenLubeck2008,PruessnerBook}.

\subsection{Phenomenology of the Oslo model}
Assume that we start with a system containing many grains. Thanks to the  Dirichlet boundary condition for $h$ to the right, the system looses  grains ($h$-particles) there until it reaches a critical slope.   This critical slope is equivalent to a  critical density of $z$-particles, which leave the system  at the left.\footnote{Note that since due to \Eq{z-def} $z$ is the derivative of $h$. Thus Neumann boundary conditions for $h$ are Dirichlet boundary conditions for $z$, and vice versa.}
We note here that sites with no $z$-particles or a single $z$-particle are always stable, sites with two $z$-particles could be stable or unstable, and particles with three or more $z$-particles are always unstable. This means that the critical slope must be between $1$ and $2$ (in fact, it equals approximately $1.8$).

An important feature of the Oslo model is that it is Abelian (commutative). In our definition we explained which topplings are invoked, but we did not mention the order in which they occur. It is not difficult to see that the final configuration, after all topplings, does not depend on this order: $z(i)$  is given by the number of incoming particles (the topplings at $i-1$ and $i+1$) and the number of outgoing particles (twice the topplings at $i$). Since a toppling at one site cannot render another site inactive, the total number of topplings at each site does not depend on the order, hence  the final configuration also does not depend on the order.

Thanks to the Abelianity of the model, we are allowed to choose freely the order of topplings. In order to compare the Oslo model with the models presented above, we choose the same type of dynamics, making each site topple with rate $1$ if $z(i)>z_{\rm c}(i)$. 

The Oslo model has been studied extensively, and impressively accurate simulations of its critical exponents are known \cite{GrassbergerDharMohanty2016}.
For the sake of this paper we wish to mention two of these exponents.
First, the roughness exponent is conjectured \cite{GrassbergerDharMohanty2016} to be $\zeta=\frac{1}{4}$, that is, $\left \langle \left[h(i)-h(j)\right]^2 \right \rangle   \sim |i-j|^{2\zeta}$. In the language of the $z$-particles, this is a hyperuniform state \cite{HexnerLevine2015}, i.e., the number of particles between $i$ and $j$ fluctuates as $|i-j|^\zeta \ll |i-j|^{1/2}$.
The second relevant exponent  is the dynamic exponent $z_\text{Oslo}$, conjectured \cite{GrassbergerDharMohanty2016} to be   $z_\text{Oslo} = \frac{10}{7}$.

\subsection{Why is  $\zeta=\frac{1}{4}$ in the Oslo model?} 
\subsubsection{Local time and global time} The quenched noise of the Oslo model may be thought of as a local time shift: $z_c$ is only updated after a toppling, while an annealed version would update $z_c$ at a fixed rate. We therefore define the \emph{local time} at a site $x$ as the number of toppling events at $x$. This   delays the topplings as compared to the models discussed in sections \ref{Particle-hole symmetric case -- the aaEW universality class}, but when the topplings occurr, their dynamics is faster, so as to ``catch up''. As a result, we expect the dynamical critial exponent $z$ to be smaller than $2$, and indeed $z_\text{Oslo} \approx \frac{10}{7}<2$.

\subsubsection{Hydrodynamic behaviour during the lifetime of an avalanche} A key observation for the Oslo model is that, during an avalanche, the local time and the global time propagate approximately at the same rate. Therefore, as long as we are interested in the evolution of an avalanche at times which are much shorter than its survival time, we may replace the quenched noise with an annealed one.
\subsubsection{Comparing the Oslo model with aaEW}
As discussed in the previous section, we consider time scales much shorter than the lifetime of an avalanche. We can therefore assume that the system evolves according to a hydrodynamic limit of the type in equation \eqref{eq:general_HL}, with an anchored boundary (at the left). In this model, the analogue of the particles in the DEP or the Toom model are the stress variables $z$.

\begin{figure*}[t]
\centerline{%
\fig{0.49}{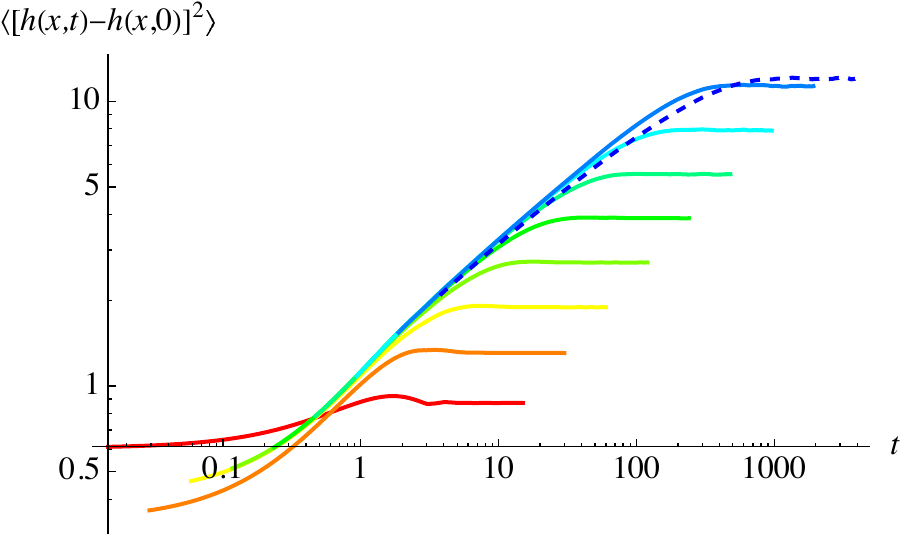}~
\fig{0.49}{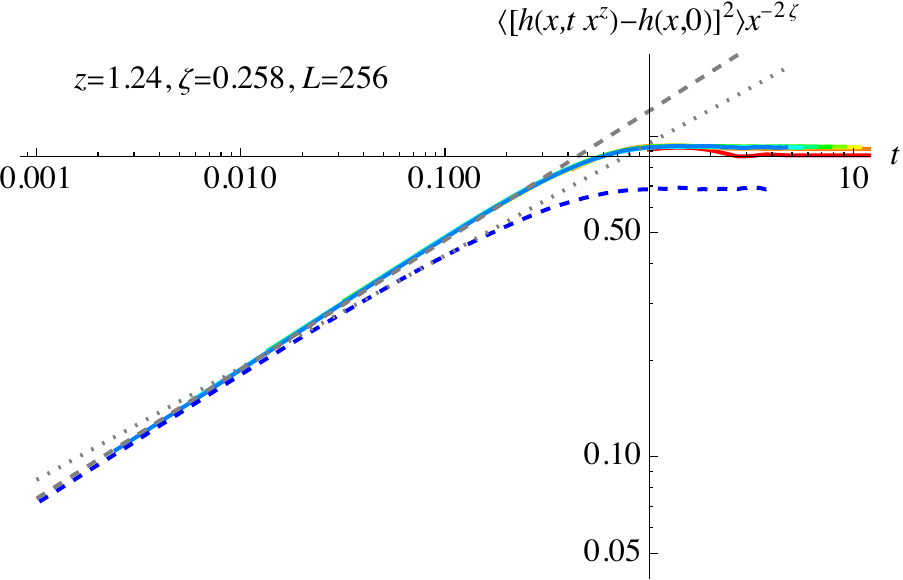}}
\centerline{\fig{0.49}{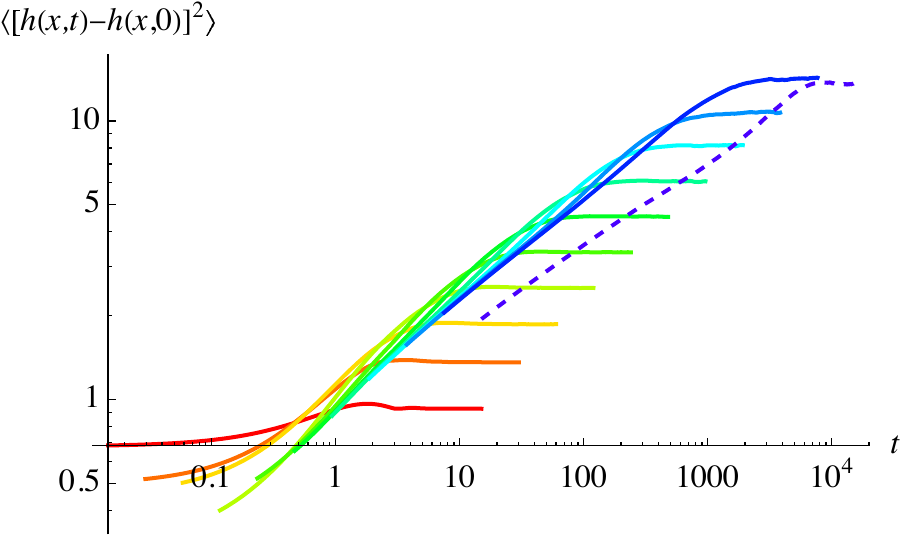}~
\fig{0.49}{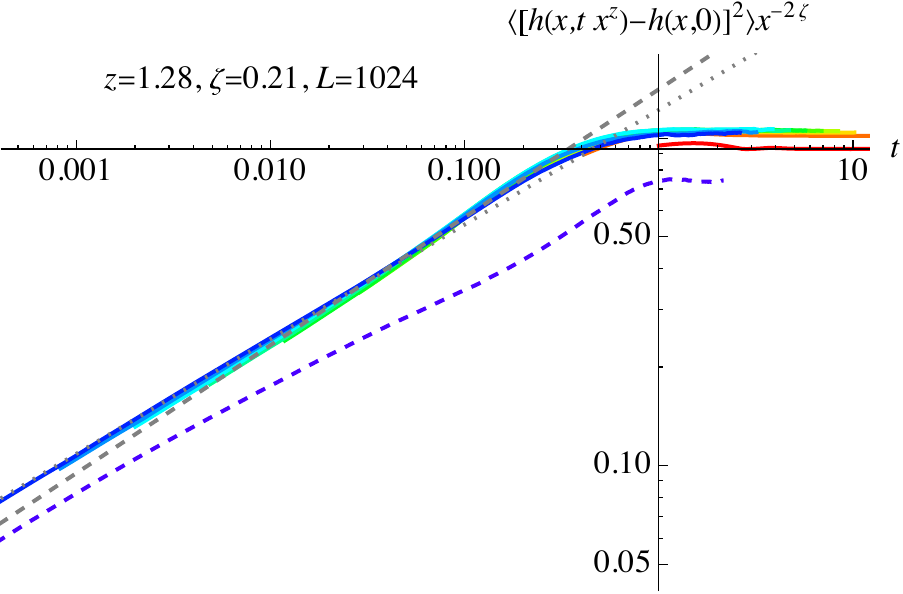}}
\caption{Top line: The Oslo model at driving rate  $r=0$, $L=256$, $x=1$, $2$, $4$, ..., $256$ (from red over yellow, green, cyan, blue to violet). The last point is at $x=L$, drawn in dashed. It   behaves differently from bulk points. The dashed gray line has power $2\zeta/z$ with best-fit values as indicated; the dotted line corresponds to the theoretical prediction $2\zeta/z=7/20$ ($\zeta=1/4$, $z=10/7$.)
Bottom line: {\em ibid} with driving rate  $r=1/5$, and $L=1024$.   See the discussion in section \ref{s:LAdiabatic versus finite advection in the Oslo model}.}
\label{f:Oslo-temp-collapse1}
\end{figure*}

Unlike the DEP or Toom's model, in the Oslo model not only the number of $z$-particles is conserved but also their center of mass, i.e., 
\begin{equation} \label{eq:oslo_conservation}
\partial_t \int_w^y h(x,t) \,\rmd x = \text{topplings at }w - \text{topplings at }y.
\end{equation}
Let us consider each of the terms in \Eq{eq:general_HL}: 
The diffusion term $D \nabla^2 h$ and the advection term $\mu \nabla h$ integrate to 
\bea
\int_w^y \rmd x \,
D   \nabla^2 h(x,t)  &=& 
D [\nabla h(y,t) - \nabla h(w,t)] , \\
\int_w^y \rmd x \, \mu\nabla h(x,t) &=& \mu[  h(y,t) -   h(w,t)],
\eea both bounded uniformly in the interval length $y-w$ (i.e.\ they remain bounded even if $y-w$ becomes large). 
The quadratic and constant terms are given by 
\bea
\int_w^y \rmd x \, 
\big[ \lambda_2 \nabla(h(x,t))^2 + c \big]  
\eea
The zero-current condition \eqref{eq:0current} implies that this term vanishes on average. However its fluctuations grow with $y-w$, so cannot be bounded uniformly. This means that $\lambda_2=0$.
The cubic term
\be
\int_w^y \rmd x \, 
\lambda_3 [\nabla h(x,t)]^3 
\ee
is also a fluctuating term which cannot be bounded uniformly, unless $\lambda_3=0$. 
In summary, the conservation of the center of mass forces $\partial_t h(x,t)$ to be a total derivatives, which is a stronger constraint than the particle-hole symmetry in Toom or the directed exclusion process\footnote{We already point out that this is  related to the fact that $h(x,t)$ can be thought of as the gradient of a field $u(x,t)$, see \Eq{h=du}.}.

The argument above tells us that, for times shorter than $L^{z_\text{Oslo}}$, the Oslo model has an aaEW behaviour. Since $z_\text{Oslo} > z_\text{aaEW}=1$, we are allowed to consider the model up to times $L^{z_\text{aaEW}}$. By then the surface reaches a roughness of $\zeta_\text{aaEW}=\frac{1}{4}$, which remains the roughness of the Oslo surface.

\subsection{Adiabatic versus finite advection in the Oslo model}
\label{s:LAdiabatic versus finite advection in the Oslo model}

In Fig.~\ref{f:Oslo-temp-collapse1}  we show the  auto-correlation function in the Oslo model, as a function of ``grain-time'' $t$: the protocol is to add one grain, and then to let the full system relax during $n=1/r$ iterations, or until all sites are stable. As a result, the time $t$ in this simulation is the number of  added  grains, and $r$ the driving rate. What we measure is the auto-correlation function in the steady state,
\be
\left< \left[ h(x,t)-h(x,0) \right]^2\right> \equiv \left< \left[ h(x,t+t')-h(x,t') \right]^2\right> .
\ee
One observes on Fig.~\ref{f:Oslo-temp-collapse1} that for adiabatic driving ($r\to 0$)
\begin{enumerate}
\item[(i)] For each $x$, it reaches a plateau after some time $\tau_x$. The larger $x$, the larger $\tau_x$.
\item[(ii)]
A scaling collapse can be achieved by the ansatz
\be
  \left< \left[ h(x,t)-h(x,0) \right]^2\right> \simeq x^{2\zeta} f_x\left(t/x^{z} \right) \quad \Longleftrightarrow \quad f_x(t) = \left< \left[ h(x,tx^z)-h(x,0) \right]^2\right>,  
\ee
where $f_x(t) \to f(t)$ for $x=1,2,4,...,L/2$.
\item[(iii)]
The best scaling collapse is achieved by $\zeta=0.258$ (close to the predicted $\zeta=1/4$), and $z=1.24$, definitely smaller than $z=10/7=1.42857$.
\item[(iv)]
The last point $x=L$ behaves differently, and its scaling function $f_L(t)$ does not collapse together with the others onto a master curve. 
While $f_x(t)\sim t^{2\zeta/z}$ with a power given by $\zeta=0.258$, and $z=1.24$, the last point has a slope approaching $2\zeta/z=7/20$.
\end{enumerate}
Let us now turn to a finite driving rate $r=1/5$, i.e.\ after adding one grain on the left, we try to topple each site 5 times. Having a finite injection rate, $t$ can both be interpreted as the number of injected grains, or time (divided by 5). 
We now observe that 
\begin{enumerate}
\item[(i)] Using the theoretically predicted values of $\zeta=1/4$ and $z=10/7$ results in a decent scaling collapse, which improves for larger $x$. 
\item[(ii)] The slope of $f(t)$ in a logarithmic scale does not seem to approach $2\zeta/z$, except for the last point. 
\item[(iii)] Things improve for larger system sizes.
\item[(iv)] Remarkably, while we studied smaller driving rates $r=1/20$ and $r=1/10$ (not shown) even at the large driving rate of $r=1/5$ the plots are almost unchanged as compared to $r=0$, indicating that we are still in a critical state.
\item[(v)] This critical state  at large driving can be described by the \emph{anchored advected Edwards Wilkinson} equation \eq{e1}.
\end{enumerate}
This confirms our findings that the roughness $\zeta=1/4$ in the Oslo model has the same origin   as in the anchored Edwards-Wilkinson equation \eq{eq:aaEW}.

\subsection{Oslo and quenched Edwards-Wilkinson}
Here we give the relation to depinning in the Edwards-Wilkinson model, of a string of size $L$ \cite{Frette1993,ChristensenCorralFretteFederJossang1996,GrassbergerDharMohanty2016,Wiese2021}. The latter has an equation of motion, 
\be
\partial_t u(i,t) = \nabla^2 u(i,t) + F(i,u(i,t)),
\ee
which we   read discretized in time $t$ and space $u$. The {\em random forces} $F(i,u)$ are   uncorrelated Gaussian random variables with 
unit variance. To connect this   to the Oslo sandpile, define
\be\label{h=du}
u(i):=  \sum_{j=i+1}^L h(i) + \# \{ \mbox{grains fallen off at the right end}\}  .
\ee
As a consequence, the discrete Laplacian
\be
\nabla^2 u(i) = z(i), 
\ee
and the random force $F(i,u)$ stems from the fact that if $u(i)\to u(i)+1$, $F(i,u)\to F(i,u+1)$ is a new random variable, which identifies   with $z_{\rm c}(i)$ in the Oslo model.  The interpretation is that of a string pulled at site $i=0$ (its left end), with Neumann boundary conditions (no force) at the right end.  
Its average profile is parabolic,  
\bea
\left< u(i) \right> \approx \frac{\left< z\right>}{2} (L-i)^2 +  \# \{ \mbox{grains fallen off at the right} \}.\nn\\
\eea
As the disorder is renewed after each displacement, it falls into the random-field universality class of depinning \cite{Wiese2021}. 
Since $\nabla u \sim h$, a  roughness exponent $\zeta=\frac{1}{4}$ for $h$ is the same as  a roughness exponent 
\be
\zeta^{\rm qEW}_{d=1} = \frac{5}{4}
\ee
for $u$. This is what we wanted to show.

\section{Conclusion}
As we have shown, 
there is a large class of models which have a roughness exponent of $\zeta=1/4$. This encompasses the Toom model, advected diffusion with an absorbing boundary, the symmetric directed exclusion process,     and the Oslo model. 
Observing that the height in the Oslo model corresponds to the slope of the height in the quenched Edwards-Wilkinson model, we showed that the roughness exponent of the latter is $\zeta_{\rm qEW}=5/4$.
This puts onto a firm basis the recent conjecture \cite{GrassbergerDharMohanty2016} that the numerically observed exponent of $\zeta_{d=1}^{\rm qEW}=1.25$ \cite{FerreroBustingorryKolton2012} is exactly $\zeta_{d=1}^{\rm qEW}=5/4$.

While these models have the same roughness exponent, their dynamical exponent $z$ can be quite different, and depend on the observable. We identified $z$ for the first three models, the {\em anchored advected universality class}, where we observed a dynamical exponent of $z=2$, $z=1$ and even $z=1/2$, depending on the obversable. 
Our correspondence to the Oslo model cannot (yet)  provide its dynamic exponent $z$, conjectured to be $z=10/7$, since the arguments use both a global and a local time, which are distinct in this case. 
Given that $z=10/7$ is again a simple fraction, we hope that an exact argument may be found as well.

\subsection*{Acknowledgements}
We are grateful to Joachim Krug for explaining how advection can convert temporal correlations into spatial ones, and for directing us to the relevant literature. 
We   thank Peter Grassberger for a useful exchange.  

%
%
%

\ifx\doi\undefined
\providecommand{\doi}[2]{\href{http://dx.doi.org/#1}{#2}}
\else
\renewcommand{\doi}[2]{\href{http://dx.doi.org/#1}{#2}}
\fi
\providecommand{\link}[2]{\href{#1}{#2}}
\providecommand{\arxiv}[1]{\href{http://arxiv.org/abs/#1}{#1}}
\providecommand{\hal}[1]{\href{https://hal.archives-ouvertes.fr/hal-#1}{hal-#1}}
\providecommand{\mrnumber}[1]{\href{https://mathscinet.ams.org/mathscinet/search/publdoc.html?pg1=MR&s1=#1&loc=fromreflist}{MR#1}}

\tableofcontents

\end{document}